\documentclass[conference]{IEEEtran}
\IEEEoverridecommandlockouts
\usepackage{cite}
\usepackage{amsmath,amssymb,amsfonts}
\usepackage{algorithmic}
\usepackage{graphicx}
\usepackage{textcomp}
\usepackage{xcolor}
\usepackage{multirow}
\usepackage{ulem}
\usepackage{subfigure}
\usepackage{booktabs}
\usepackage{fancyhdr}

\def\BibTeX{{\rm B\kern-.05em{\sc i\kern-.025em b}\kern-.08em
    T\kern-.1667em\lower.7ex\hbox{E}\kern-.125emX}}
    
\begin{document}

\title{BAFLineDP: Code Bilinear Attention Fusion Framework for Line-Level Defect Prediction}

\makeatletter
\newcommand{\linebreakand}{%
    \end{@IEEEauthorhalign}
    \hfill\mbox{}\par
    \mbox{}\hfill\begin{@IEEEauthorhalign}
}
\makeatother

\author{\IEEEauthorblockN{Shaojian Qiu}
\IEEEauthorblockA{\textit{College of Mathematics and Informatics} \\
\textit{South China Agricultural University}\\
Guangzhou, China \\
qiushaojian@scau.edu.cn}
\and
\IEEEauthorblockN{Huihao Huang}
\IEEEauthorblockA{\textit{College of Mathematics and Informatics} \\
\textit{South China Agricultural University}\\
Guangzhou, China \\
huanghuihao@insoft-lab.com}
\and
\IEEEauthorblockN{Jianxiang Luo}
\IEEEauthorblockA{\textit{College of Mathematics and Informatics} \\
\textit{South China Agricultural University}\\
Guangzhou, China \\
luojianxiang@insoft-lab.com}
\linebreakand
\IEEEauthorblockN{Yingjie Kuang}
\IEEEauthorblockA{\textit{College of Mathematics and Informatics} \\
\textit{South China Agricultural University}\\
Guangzhou, China \\
kuangyj@scau.edu.cn}
\and
\IEEEauthorblockN{Haoyu Luo$^*$}
\IEEEauthorblockA{\textit{College of Mathematics and Informatics} \\
\textit{South China Agricultural University}\\
Guangzhou, China \\
haoyuluo@scau.edu.cn}
}

\maketitle

\thispagestyle{fancy}
\lfoot{\rule[-\dp\strutbox]{0pt}{\baselineskip}} 
\lhead{}
\lfoot{\footnotesize{Accepted at the Int. Conf on Software Analysis, Evolution and Reengineering SANER 2024, March 12–15, 2024, Rovaniemi, Finland. \\ ©2024 IEEE. Personal use of this material is permitted. Permission from IEEE must be obtained for all other uses, in any current or future media, including reprinting/republishing this material for advertising or promotional purposes, creating new collective works, for resale or redistribution to servers or lists, or reuse of any copyrighted component of this work in other works.}}
\cfoot{}
\rfoot{}
\renewcommand{\headrulewidth}{0pt}

\begin{abstract}
Software defect prediction aims to identify defect-prone code, aiding developers in optimizing testing resource allocation. Most defect prediction approaches primarily focus on coarse-grained, file-level defect prediction, which fails to provide developers with the precision required to locate defective code. Recently, some researchers have proposed fine-grained, line-level defect prediction methods. However, most of these approaches lack an in-depth consideration of the contextual semantics of code lines and neglect the local interaction information among code lines. To address the above issues, this paper presents a line-level defect prediction method grounded in a code bilinear attention fusion framework (BAFLineDP). This method discerns defective code files and lines by integrating source code line semantics, line-level context, and local interaction information between code lines and line-level context. Through an extensive analysis involving within- and cross-project defect prediction across 9 distinct projects encompassing 32 releases, our results demonstrate that BAFLineDP outperforms current advanced file-level and line-level defect prediction approaches.
\end{abstract}


\section{Introduction}
As the scale and complexity of modern software continue to increase, software development and maintenance are becoming arduous. Software defects not resolved in time will inevitably affect the software quality~\cite{meng2021semi}. Defect prediction can help the quality assurance teams discover potential defects during the development process, contributing to optimizing the allocation of limited testing resources. To facilitate this, researchers~\cite{thota2020survey} mine data from software historical repositories, construct code metrics and representations, and train prediction models to pinpoint defect-prone areas in software.

In recent years, researchers have proposed defect prediction methods at various granularity levels, such as package level~\cite{kamei2010revisiting}, component level~\cite{thongtanunam2016revisiting}, module level~\cite{kamei2007effects}, file level~\cite{kamei2010revisiting, mende2010effort} and method level~\cite{hata2012bug}. Current research indicates that a more granular approach to defect prediction in software testing yields greater cost-effectiveness in code inspection~\cite{hata2012bug, kamei2010revisiting, pascarella2019fine}. Kamei et al.~\cite{kamei2010revisiting} demonstrated that defect prediction at the file level is more effective than at the package level. Similarly, Hata et al.~\cite{hata2012bug} pointed out that method-level defect prediction offers superior cost-efficiency over file-level defect prediction. Although these methods have been verified through empirical studies, they are still coarse-grained predictions and are difficult to help developers in actual defect-finding tasks. For example, file-level defect prediction can only show developers the defect-prone files. Developers still have to spend extra effort walking through files and identifying defective code lines, resulting in inefficient reviews. Therefore, studying more fine-grained line-level defect prediction is necessary to improve the efficiency of code review work.

Currently, a handful of researchers are attempting to identify the defective lines of code. Wattanakriengkrai et al.~\cite{wattanakriengkrai2020predicting} adopt a model-agnostic technique to discern code tokens with defective risks and classify the code containing the risk tokens as defective lines. Based on ensemble learning and attention mechanism, Zhang et al.~\cite{zhang2020software} used abstract syntax trees to build a  model to predict method-level defects and locate suspicious code lines. Pornprasit et al.~\cite{pornprasit2022deeplinedp} proposed a method called DeepLineDP, which utilizes a hierarchical Bi-GRU~\cite{zhu2020speech} network to extract code features to predict file-level defects and identify defective code lines through the token weights.

While the aforementioned approaches excel in line-level defect prediction tasks, they mostly neglect the contextual semantics of code lines and the local interaction information between them. Specifically, these approaches mainly rely on token-level attention within code lines to gauge their potential for defects. However, the computation of these token-level attentions is constrained to the information contained solely within the confines of the individual code line. It often fails to fully encompass the broader contextual information pertaining to the code line. Furthermore, software defects may manifest when multiple lines of code interact locally, leading to issues such as null pointer exceptions. Token attention-based methodologies may encounter challenges in effectively predicting such types of flaws. Consequently, considerable room remains for enhancement within the domain of existing line-level defect prediction methods.

To elucidate the problem of missing code line context and local interaction information between code lines and line-level context, we present an illustrative abridged case in Figure~\ref{motivation-case}. In this case, there is a potential NullPointerException. Line 6 of the code directly calls $name.length()$ to get the length of the string without checking whether the variable is null. If $jo.get("name")$ returns null, a NullPointerException will be thrown. The above-mentioned line-level defect prediction method mainly focuses on defects caused by tokens within code lines, and cannot mine the local contextual association between the 3-4 lines and the 6th line of code. Early warning of the defect is thus missed.

\begin{figure}
    \centerline{
    \includegraphics[width=0.8\linewidth]{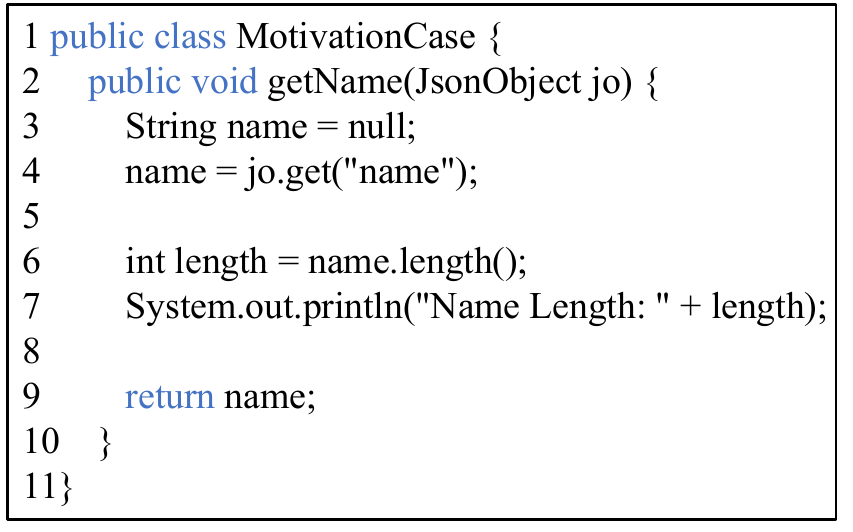}}
    \caption{A motivation case of missing line context and local interaction between code lines and line-level context.}
    \label{motivation-case}
\end{figure}

To solve the above problems, this paper proposes a code bilinear attention fusion framework for line-level defect prediction (BAFLineDP). Specifically, BAFLineDP first uses a pre-trained CodeBERT model to parse the code file into a line embedding matrix. Subsequently, we adopt Bi-GRU to extract contextual features between code lines. Then, a bilinear attention fusion network is employed to mine the interactive representation of each code line and corresponding line-level context. Finally, we use code representations generated by BAFLineDP to improve prediction performance on defect prediction tasks within and across projects. The empirical results demonstrate that the BAFLineDP approach is better than other advanced defect prediction methods in the AUC, MCC, and BA metrics of file-level defect prediction and the Recall@Top20\%LOC and Effort@Top20\%Recall indicators of line-level defect prediction.

The main contributions of this paper are as follows:

\begin{enumerate} 
\setlength{\itemsep}{-2ex}  
\setlength{\parskip}{0ex} 
\setlength{\parsep}{0ex}

\item[-] This paper introduces a bilinear attention fusion mechanism into the line-level defect prediction framework, taking into full account the global line-level context and local interactions of code lines to deeply mine the representations of code features. This method addresses the issue in existing line-level defect prediction approaches, which lack a comprehensive consideration of both global and local information at the code line level.\hfil\break

\item[-] This paper conducts experiments on multiple open-source benchmark projects and compares existing advanced methods. Experimental results show that the proposed BAFLineDP method achieves considerable prediction performance in defect prediction tasks within and across projects.
\end{enumerate}

\section{Related Work}
Software defect prediction is one of the important research directions in the field of intelligent software engineering. Its purpose is to help software quality assurance teams rationally allocate limited testing resources to improve the quality of software products. Researchers in this field have proposed defect prediction methods for various granularity, such as package level~\cite{kamei2010revisiting}, component level~\cite{thongtanunam2016revisiting}, module level~\cite{kamei2007effects}, file level~\cite{kamei2010revisiting, mende2010effort} and method level~\cite{hata2012bug}. Currently, most mainstream research focuses on file-level defect prediction.

File-level defect prediction methods are mainly used to predict whether code files have defect tendencies. These methods are usually based on the static attributes~\cite{matloob2021software} of code (e.g., Halstead features, McCabe features, CK features, and MOOD features), change attributes~\cite{rhmann2020software} (e.g., process characteristics organizational structure, code ownership, number of code revisions, change entropy, and developer characteristics) and structural-semantic features extracted by deep neural networks~\cite{qiao2020deep, yedida2021value, chen2020software}. Most existing file-level defect prediction methods perform model training and evaluation within the same project. However, in actual application scenarios, it is not easy to obtain enough training samples for new projects. Therefore, cross-project file-level defect prediction approaches~\cite {deng2020suitable, sun2021cfps, xing2022cross} have been proposed to address data scarcity in target projects. These methods leverage high-quality datasets collected from other projects to construct a defect prediction model for the target project. Whether in defect prediction scenarios within or across projects, file-level defect prediction has been confirmed by most studies to achieve notable performance. However, file-level defect prediction is still a coarse-grained prediction, which makes it challenging to help developers find defects in actual application scenarios. The main reason is that file-level defect prediction can only predict the defect tendency of code files. Developers still need to spend additional time traversing files and identifying defective lines, which affects their efficiency in locating and fixing defects.

\begin{figure*}[t]
    \centering
    \includegraphics[width=1.0\textwidth]{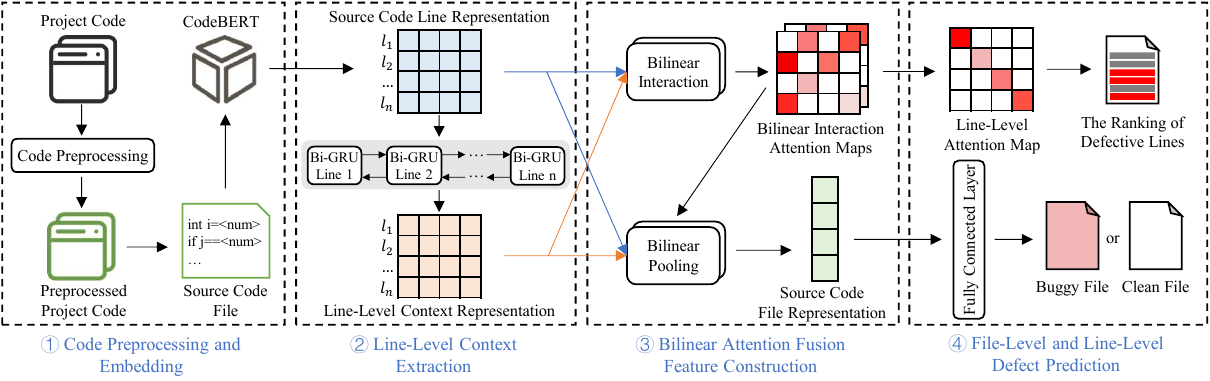}
    \caption{An overview of our BAFLineDP framework.}
    \label{fig:enter-label}
\end{figure*}

Compared with file-level defect prediction, line-level defect prediction can discover more fine-grained defects and help developers locate defects more accurately and efficiently. A survey by Wan et al.~\cite{wan2018perceptions} found that software practitioners are more inclined to use fine-grained defect prediction to assist code review work. In recent years, some researchers have tried to utilize various approaches to predict defective code lines~\cite{majd2020sldeep, pornprasit2021jitline, wang2016bugram, wattanakriengkrai2020predicting}. One of the simplest methods is to use static analysis tools to identify defective lines of code based on predefined rules~\cite{johnson2013don}. However, static analysis tools may generate many false positive warnings unrelated to defects. Majd et al.~\cite{majd2020sldeep} designed 32 statement-level features based on C/C++ and used a long-term memory network to build a statement-level defect prediction model. However, these statement-level features require manual knowledge extraction by experts and can only capture the static attributes of code statements. Ray et al.~\cite{ray2016naturalness} and Wang et al.~\cite{wang2016bugram} employed the n-gram model to capture surrounding code tokens and discern defective lines by identifying unnatural tokens. However, the n-gram model can only capture surrounding tokens with a limited length, which is insufficient to generate effective code context features. Recently, Pornprasit et al.~\cite{pornprasit2022deeplinedp} proposed a line-level defect prediction method called DeepLineDP, which automatically extracts semantic features from tokens in lines to predict file-level defects and identify defect-prone code lines through the attention of code tokens. However, DeepLineDP's calculation of token attention only focuses on the information of the current code line, and it is difficult to capture the contextual semantics and local interactions between code lines effectively.

Although the above approaches are better than most existing defect prediction methods~\cite{johnson2013don, majd2020sldeep, ray2016naturalness} in line-level defect prediction tasks, it suffers from the problem of missing a comprehensive consideration of both global and local information at the code line level. The methodology presented in this paper comprehensively considers the line-level context and local interactions among code lines, enabling deep mining of code feature representations.

\section{Methodology}
In this section, we present BAFLineDP, a line-level defect prediction approach based on a bilinear attention fusion framework. Figure 2 provides a holistic overview of our BAFLineDP framework, which encompasses four steps: (1) Code Preprocessing and Embedding; (2) Line-Level Context Extraction; (3) Bilinear Attention Fusion Feature Construction; and (4) File-Level and Line-Level Defect Prediction.

\begin{figure*}[t]
    \centering
    \includegraphics[width=0.7\textwidth]{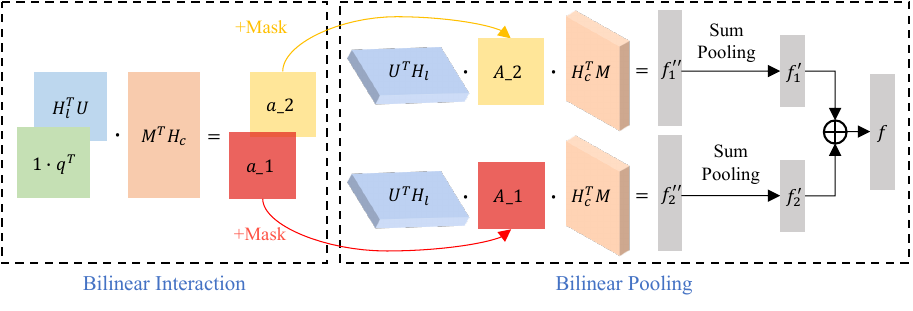}
    \caption{The process of feature construction with bilinear attention fusion.}
    \label{fig:enter-label}
\end{figure*}

\subsection{Code Preprocessing and Embedding}
Code preprocessing~\cite{singh2020investigating} plays a pivotal role within the realm of deep learning. It serves as a critical step in enhancing the robustness and stability of models by standardizing and normalizing the source code. By eliminating redundant information that holds no relevance to code logic, the impact of noise on the model is reduced, allowing the model to extract code semantics and syntactic features more effectively. Therefore, we performed code preprocessing for each code file. Based on the investigation by Rahman et al.~\cite{rahman2019natural}, we systematically removed all special characters (e.g., {}(),.:;'!"(space)) since these special characters introduce undesirable noise to the prediction model. Simultaneously, blank lines were also removed because of no substantive contribution to the code's behavior~\cite{hoang2020cc2vec}. Eliminating such extraneous information allows the model to maintain its focus on learning the actual content of the code, free from interference by irrelevant factors. In addition, to enhance the model's generalization capability, we introduce the generic tokens $\langle$str$\rangle$ and $\langle$num$\rangle$ to replace constant strings and numeric representations within the source code. This strategic substitution ensures the model does not generate independent representations for these elements, rendering it more adaptable to diverse scenarios and usage cases.

In our pursuit of capturing precise semantics related to defective code, we utilize the pre-trained CodeBERT model to encode code lines. CodeBERT~\cite{feng2020codebert} is a bimodal pre-trained language model that leverages a multi-layer bidirectional Transformer as the architecture to capture the meaningful semantic relationships among diverse tokens, contributing to construct an adequate representation of the source code line. The incorporation of the pre-trained CodeBERT model holds the promise of a deeper understanding of code line semantics and the accurate capture of features associated with defects. Furthermore, to maintain the structural information of the source code, we adopt a sequential structure to depict the organization of code files. In this configuration, each code file is expressed as a sequence of code lines, denoted as $\langle l_1,l_2,\cdots,l_n \rangle$. For each code line, the pre-trained CodeBERT model will generate the corresponding embedded representation. 

\subsection{Line-Level Context Extraction}
In this step, we employ the Bi-GRU network to extract line-level contextual features, thereby capturing the global information pertaining to code lines. The Bi-GRU~\cite{zhu2020speech} is founded upon a bidirectional recurrent neural network architecture, which affords it the capability to concurrently consider both the preceding and succeeding information within sequences of source code lines. The bidirectional mode ensures the effective modeling of the contextual associations among code lines, culminating in a richer representation of their interdependencies.

Given a sequence of code lines, denoted as $V_l=[v_{l1},v_{l2},\cdots,v_{ln}]$, where $v_{li} \in R^{1\times d}$ represents the vector representation of code line $l_i$, $i \in [1,n]$. These representations are encoded by the pre-trained CodeBERT model, where $d$ signifies the output dimension of CodeBERT. Bi-GRU comprises both a forward GRU denoted as $\overrightarrow{h}_i=\overrightarrow{GRU}(v_{li})$, $i\in [1,|n|]$ and a backward GRU denoted as $\overleftarrow{h}_i=\overleftarrow{GRU}(v_{li})$, $i\in [|n|,1]$. Through the concatenation of the two hidden states, $\overrightarrow{h}_i$ and $\overleftarrow{h}_i$, produced by the forward and backward GRU components, we arrive at the final hidden representation of the given $v_{li}$. In other words, the line-level context vector representation can be expressed as $v_{ci}=[\overrightarrow{h}_i \oplus \overleftarrow{h}_i]$. Consequently, the line-level context sequence is represented as $V_c=[v_{c1},v_{c2},\cdots,v_{cn}]$, where $v_{ci} \in R^{1\times d^{'}}$ characterizes the line-level context vector representation of code line $l_i$, and $d^{'}$ signifies the output dimension of Bi-GRU.

\subsection{Bilinear Attention Fusion Feature Construction}
Source code is inherently complex and structured~\cite{le2020deep}, encompassing not only overall structure and semantic features but also local intricacies and contextual dependencies. Merely considering line-level context information is insufficient for a comprehensive understanding of defective code lines. Hence, to address this limitation, we introduce a bilinear attention mechanism~\cite{kim2018bilinear} to build a code fusion network, called BAFN. BAFN is designed to amalgamate global and local information by capturing the bilinear interaction attention weights between code lines and their respective line-level contextual information to construct defect code features. The structure of BAFN, as depicted in Figure 3, primarily includes two essential modules: the bilinear interaction module and the bilinear pooling module.

\uline{Bilinear Interaction Module.} Assume that the code line matrix denoted as $H_l\in R^{\theta_l\times d}$ and the line-level context matrix denoted as $H_c\in R^{\theta_l\times d^{'}}$ are constructed by splicing the vectors from the code line sequence $V_l$ and the line-level context sequence $V_c$, respectively. $\theta_l$ represents the total number of code lines. We construct the bilinear interaction matrix $A\in R^{\theta_l\times \theta_l}$ through the source code line matrix $H_l$ and the line-level context matrix $H_c$. The calculation formula is as follows:

\begin{equation}
    \label{eq:equa1}
    A=((1\cdot q^{T})\circ \sigma (H_{l}^{T}U))\cdot \sigma (M^{T}H_c)
\end{equation} where $U\in R^{d\times k}$ and $M\in R^{d^{'}\times k}$ are two learnable linear transformation matrices, which are also weight matrices. $k$ represents the dimension of the linear transformation, $q\in R^{1\times k}$ is a learnable weight vector. The symbol $\circ$ denotes the Hadamard product operation~\cite{arvind2022fast}, and $\sigma (\cdot)$ signifies the ReLU activation function~\cite{schmidt2020nonparametric}. Each element within the bilinear interaction attention matrix $A$ encapsulates the extent of interaction between a source code line and its corresponding line-level context, further presenting the local associations between potential defects and code lines. In addition, to prevent the model from excessively focusing on extraneous noise information, we incorporate the mask into the interaction attention map $A$ to enhance the overall model performance.

\begin{table*}[htbp] 
	\center
    \setlength{\tabcolsep}{5pt}
	\renewcommand\arraystretch{1.2}
	\caption{An Overview of the Experimental Datasets}
	\begin{tabular}{c c c c c c c} 
		\toprule
		\textbf{Project}&\textbf{Description}&\textbf{Release Version} &\textbf{\#File} & \textbf{\#LOC} & \textbf{\%Defective Files} & \textbf{\%Defective Lines} \\ 
		\midrule
		ActiveMQ & Messaging and integration patterns & 5.0.0, 5.1.0, 5.2.0, 5.3.0, 5.8.0  & 1,884-3,420  & 142k-299k & 2\%-7\%  & 0.08\%-0.44\%  \\
		Camel    & Enterprise integration framework   & 1.4.0, 2.9.0, 2.10.0, 2.11.0       & 1,515-8,846  & 75k-485k  & 2\%-8\%  & 0.09\%-0.24\%  \\
		Derby    & Relational database                & 10.2.1.6, 10.3.1.4, 10.5.1.1       & 1,963-2,705  & 412k-533k & 6\%-28\% & 0.10\%-0.63\% \\
		Groovy   & Java-syntax-compatible OOP         & 1.5.7, 1.6.0.Beta1, 1.6.0.Beta2    & 757-884      & 74k-93k   & 2\%-4\%  & 0.10\%-0.17\% \\
		HBase    & Distributed scalable data store    & 0.94.0, 0.95.0, 0.95.2             & 1,059-1,834  & 246k-537k & 7\%-11\% & 0.17\%-1.02\% \\
		Hive     & Data warehouse system for hadoop   & 0.9.0, 0.10.0, 0.12.0              & 1,416-2,662  & 290k-567k & 6\%-19\% & 0.31\%-2.90\% \\
		JRuby    & Ruby programming lang for JVM      & 1.1, 1.4, 1.5, 1.7                 & 731-1,614    & 106k-240k & 2\%-13\% & 0.03\%-0.09\% \\
		Lucene   & Text search engine library         & 2.3.0, 2.9.0, 3.0.0, 3.1.0         & 805-2,806    & 101k-342k & 2\%-8\%  & 0.07\%-0.39\% \\
		Wicket   & Web application framework          & 1.3.0.beta1, 1.3.0.beta2, 1.5.3    & 1,672-2,578  & 106k-165k & 2\%-16\% & 0.05\%-0.46\% \\
		\bottomrule
	\end{tabular}
	\label{table1}
\end{table*}

\uline{Bilinear Pooling Module.} To efficiently amalgamate global and local information from the code lines and create a high-level representation of defect code, denoted as $f^{''}\in R^{1\times k}$, we employ a bilinear pooling module. This module is instrumental in extracting important features from the code. The calculation formula for $f^{''}$ is as follows:

\begin{equation}
    \label{eq:equa2}
    f^{''}=(H_l^{T}U)^{T}\cdot A\cdot (M^{T}H_c)
\end{equation}

Notably, there are no new learnable parameters introduced in this layer. Instead, the bilinear pooling module and the bilinear interaction module share the weight matrices $U$ and $M$, thereby diminishing the number of model parameters. Furthermore, we add a sum pooling operation, grounded in the fused feature representation, to obtain a more compact defect code feature representation, denoted as $f^{'}\in R^{1\times k/s}$. The calculation formula for $f^{'}$ is as follows:

\begin{equation}
    \label{eq:equa3}
    f^{'}=SumPool(f^{''},s)
\end{equation} where the $SumPool(\cdot )$ function~\cite{gholamalinezhad2020pooling} is a one-dimensional non-overlapping sum pooling operation with stride $s$, which reduces the dimensionality of $f^{''}$ from $k$ to $k/s$, allowing for the extraction of crucial features. For the pooling operation, we set the stride $s=3$.

Further, we extend the single-head bilinear interaction attention matrix into a double-head form. The final defect code fusion feature vector $f\in R^{1\times k/s}$ is a combination of two heads: $f_1^{'}\in R^{1\times k/s}$ and $f_2^{'}\in R^{1\times k/s}$. The calculation formula for $f$ is as follows:

\begin{equation}
    \label{eq:equa4}
    f=f_1^{'}\oplus f_2^{'}
\end{equation}

Since the shared utilization of weight matrices $U$ and $M$ across the two modules, introducing an additional header only requires adding a new weight vector $q$, which is parameter-efficient. Through our experiments, we find that incorporating a double-head interaction mechanism yields better performance when compared to the single-head interaction approach.

\subsection{File-Level and Line-Level Defect Prediction}
Via the BAFN, we derive a high-level vector representation $f$ of the source code file, which serves as a feature for performing file-level defect prediction. We employ a single-layer fully connected network as the prediction layer and utilize the source code file vector as input, leading to the generation of the defect prediction probability, denoted as $p$. The calculation formula is provided below:

\begin{equation}
    \label{eq:equa5}
    p=Sigmoid(W_0f+b_0)
\end{equation} where $W_0$ and $b_0$ are learnable weight matrices and bias values, and the $Sigmoid(\cdot )$ function~\cite{pratiwi2020sigmoid} serves to map the prediction score into the $(0,1)$ range.

To identify defect-prone lines of code, we leverage the diagonal elements within the bilinear interaction attention maps to rank the risk associated with each source code line. We first consolidate the bilinear interaction attention maps from a double-head form into a single-head form. Then, we extract the diagonal elements within the consolidated map to assemble the line-level attention map. The attention score on the diagonal of this map is utilized as the final risk coefficient for each code line, which enables the sorting of all the code lines within the source code file. This sorting process allows us to pinpoint the most defect-prone code lines, ultimately achieving line-level defect prediction.

\section{Experiment Design and Results}
In this section, we present the research design and experimental results, including the experiment datasets, evaluation metrics, baseline methods, experimental details, and analysis of results for each research question.

\subsection{Experiment Datasets}
In our study, we utilize the line-level defect prediction datasets collected by Wattanakriengkrai et al.~\cite{wattanakriengkrai2020predicting}, encompassing 9 open-source projects and 32 software releases. We downloaded each project version from official websites for analysis, as detailed in Table~\ref{table1}. This table includes project names, descriptions, release versions, file counts, line counts, file defect rates, and line defect rates. The datasets feature projects with a wide range of sizes and defect rates, with file counts varying from 731 to 8846, and line counts from 74k to 567k. File defect rates span 2\% to 28\%, and line defect rates range from 0.03\% to 2.90\%. For detailed steps on collecting file-level and line-level ground truth, please refer to~\cite{pornprasit2022deeplinedp}. Our experimental approach involves training on the first release of each project, validation on the second, and testing on subsequent versions. This results in 14 distinct tasks for within-project defect prediction (WPDP) and 112 for cross-project defect prediction (CPDP).

\subsection{Evaluation Metrics}
We adopt the classification metrics Area Under Curve (AUC)~\cite{bradley1997use}, Balanced Accuracy (BA)~\cite{velez2007balanced}, and Matthews Correlation Coefficient (MCC)~\cite{chicco2021matthews} to evaluate the file-level defect prediction task, and the workload-aware metrics Recall@Top20\%LOC~\cite{pornprasit2022deeplinedp} and Effort@Top20\%Recall~\cite{pornprasit2022deeplinedp} to evaluate the line-level defect prediction task.

\textbf{AUC} measures the performance of binary classification models. The AUC metric is unaffected by class distribution and, therefore, has better evaluation performance in the case of class imbalance.

\textbf{BA} measures the average ratio of true positive and true negative. The highly balanced accuracy shows that the method can accurately predict defective and clean instances.

\textbf{MCC} is used to evaluate the accuracy and stability of model predictions. MCC ranges from -1 to 1, where 1 means the predictions are perfect, 0 means the predictions are irrelevant to the actual outcome, and -1 means the predictions are completely opposite.

\textbf{Recall@Top20\%LOC} is a metric utilized to gauge the efficacy of identifying defective lines within the top 20\% LOC in a software release. A high value of Recall@Top20\%LOC signifies the method's proficiency in uncovering numerous actual defective lines with limited effort, whereas a low value of Recall@Top20\%LOC indicates that developers need to spend more effort to detect defective lines.

\textbf{Effort@Top20\%Recall} is a metric assessing the amount of effort (i.e., LOC) required to identify the top 20\% of actual defective lines in a software release. A low Effort@Top20\%Recall value suggests minimal effort is required by developers to pinpoint the top 20\% of actual defective lines, while a high Effort@Top20\%Recall value implies a greater effort is needed for the same task.

\subsection{Baseline Methods}
To evaluate the performance of our BAFLineDP approach, we compared the current advanced defect prediction methods. Below, we present four file-level defect prediction methods~\cite{wang2018deep, li2017software, hata2010fault, dam2018automatic}:

\begin{itemize}
    \item DBN, a deep belief network, automatically extracts semantic and structural code features to predict defective code files. Adhering to the experimental setup outlined by Wang et al.~\cite{wang2018deep}, our configuration comprises a batch size of 32, 10 hidden layers with 100 nodes each, an embedding dimension of 50, a learning rate of 0.01, and 200 training epochs for each hidden layer.
    \item CNN is a convolutional neural network that automatically learns code semantics and context to predict defective code files. Aligning with the experimental setup of Li et al.~\cite{li2017software}, our configuration includes a batch size of 32, an embedding dimension of 50, a filter length of 5, 100 filters, a learning rate of 0.001, and 10 training epochs.
    \item BoW (also known as Bag-of-Words), which leverages code token frequencies as features for source code files, is employed to predict defective code files. We align with the experimental deployment of Hata et al.~\cite{hata2010fault}, incorporating the SMOTE technique to address the class imbalance within training data. A logistic regression classifier is trained on code token frequency to construct the BoW model.
    \item Bi-LSTM, a bidirectional long short-term memory network, is utilized to incorporate both past and future code information, enabling the automatic learning of semantic and syntactic features for predicting defective code files. We follow the experimental configuration introduced by Dam et al.~\cite{dam2018automatic}, featuring a batch size of 16, an embedding dimension of 50, 64 nodes in the hidden layer, a learning rate of 0.001, and a training epoch of 50.
\end{itemize}

Subsequently, we introduce three advanced line-level defect prediction approaches~\cite{pornprasit2022deeplinedp, aftandilian2012building, hellendoorn2017deep}:

\begin{itemize}
    \item DeepLineDP, the state-of-the-art line-level defect prediction method, leverages a hierarchical Bi-GRU network to automatically capture context from surrounding tokens and code lines, enabling the extraction of code semantic features for file-level defect prediction. In addition, it incorporates a token-level attention layer to identify defective lines based on key tokens instrumental in file-level defect prediction. Our configuration, in line with the experimental setup of Pornprasit et al.~\cite{pornprasit2022deeplinedp}, involves a batch size of 32, a learning rate of 0.001, an embedding dimension of 50, a single hidden layer with 64 nodes, a dropout ratio of 0.2, and 10 training epochs.
    \item ErrorProne, an open-source tool by Google, leverages the Java compiler for static analysis and error detection in Java code, utilizing a set of error-prone rules to identify and rectify errors. By emulating developers' sequential reading, ErrorProne analyzes code files top-down to pinpoint potentially defective lines. Adhering to the usage guide by Aftandilian et al.~\cite{aftandilian2012building}, we invoke the ErrorProne plug-in within a Java8 compilation environment for line-level defect prediction on the test code files.
    \item N-gram is employed to assess the unnaturalness of code tokens by computing the entropy score of each token. Research conducted by Ray et al.~\cite{ray2016naturalness} reveals that defect-prone codes tend to exhibit higher entropy values, indicating increased unnaturalness. Therefore, N-gram has predictive utility in identifying defective code lines. Following the experimental setup outlined by Hellendoorn et al.~\cite{hellendoorn2017deep}, we train a cache-based N-gram model using clean code files and then perform line-level defect prediction on defective code files. By calculating and ranking the average entropy score of tokens within each line, the most defect-prone code lines could be found.
\end{itemize}

It should be noted that DeepLineDP is categorized as a line-level defect prediction method due to its hierarchical approach to constructing code features, which incorporates information from surrounding code lines. This design enhances the method's accuracy in identifying line-level defects, exhibiting state-of-the-art (SOTA) performance. However, we also consider involving DeepLineDP in file-level defect prediction for a comprehensive comparison with our BAFLineDP.

\subsection{Experimental Details}
Our BAFLineDP model is implemented using PyTorch, a deep learning framework, and executed on a server equipped with an NVIDIA RTX 3090 GPU with 24 GB memory. 

Regarding the experimental strategy, we employ binary cross-entropy as the loss function, utilize the Adam optimizer~\cite{kingma2014adam}, and incorporate random dropout (rate of 0.2) and normalization to prevent model overfitting. Additionally, we address dataset class imbalance through weighted loss.

In terms of experimental parameters, we set the batch size to 16 and 
use the learning rate of 0.001. The model is trained at 10 epochs. The maximum input length of the pre-trained CodeBERT model is 75. The Bi-GRU network is configured with one hidden layer, while the number of nodes in the hidden layer is set to 64. The output dimension for the BAFN is designated as 256. Furthermore, the convolution kernel size of the bilinear pooling layer is 3.

About evaluation models, we consider the training model that attains the highest AUC value on the validation set as the ultimate evaluation model to perform both file-level and line-level defect prediction tasks.

\subsection{Statistical Test}
Given the potential for certain datasets to yield models that exhibit significant performance discrepancies, we employ the Scott-Knott Effect Size Difference (Scott-Knott ESD) test~\cite{herbold2017comments} to compare the efficiency of different methods in our experiments. The Scott-Knott ESD test, a mean comparison technique, uses hierarchical clustering to categorize measurements (e.g., AUC) into statistically distinct groups with non-negligible effect size differences. It comprises two steps: (1) correcting the non-normal distribution of the input dataset and (2) merging any two statistically distinct combinations with negligible effect sizes into a single group. The rankings generated by the Scott-Knott ESD test ensure that (1) the differences in distribution magnitudes within each ranking category are negligible and (2) the differences in distribution magnitudes between ranks are not negligible. For a detailed description of the Scott-Knott ESD test, please refer to~\cite{herbold2017comments}.

\subsection{Research Questions and Analysis}
To assess the efficacy of our BAFLineDP approach, we conduct both file-level and line-level defect prediction in WPDP and CPDP scenarios, respectively, and compare with current advanced defect prediction techniques. Below, we present the approaches and results of two research questions (RQs).

\textbf{\textit{(RQ1) What is the performance and cost-effectiveness of BAFLineDP in WPDP scenario?}}

\uline{Motivation.} WPDP aims to identify and rectify software defects in their early stages, optimizing Software Quality Assurance (SQA) resource allocation and enhancing software quality. File-level and line-level defect prediction are techniques employed to realize these overarching objectives. Existing advanced defect prediction methodologies predominantly center on the file level, with limited attention to line-level approaches. Recently, DeepLineDP, which automatically learns the code hierarchical structure and considers code context, has been proposed to achieve SOTA performance in both file-level and line-level defect prediction within the WPDP scenario. However, this method disregards the importance of the context and local interaction information for code lines within line-level defect prediction. Therefore, we investigate whether BAFLineDP outperforms advanced file-level and line-level defect prediction methods within the WPDP scenario while offering superior cost-effectiveness.

\uline{Approach.} To answer this question, we selected 14 training-validation-testing task combinations within the WPDP scenario. For instance, training and validation were conducted on datasets such as Hive-0.9.0 and Hive-0.10.0, followed by testing on Hive-0.12.0. Subsequently, we assess the performance of the BAFLineDP method by conducting evaluations in the context of both file-level and line-level defect prediction. In order to provide a comprehensive analysis, we compare BAFLineDP with four advanced file-level defect prediction approaches~\cite{wang2018deep, li2017software, hata2010fault, dam2018automatic}, namely DBN, CNN, BoW, and Bi-LSTM. Additionally, we contrast BAFLineDP with three advanced line-level defect prediction methods~\cite{pornprasit2022deeplinedp, aftandilian2012building, hellendoorn2017deep}, which include DeepLineDP, ErrorProne, and N-gram.

To evaluate the performance of these methods, we employ three traditional evaluation metrics (i.e., AUC, MCC, and BA) for file-level defect prediction and two effort-aware evaluation metrics (i.e., Recall@Top20\%LOC and Effort@Top20\%Recall) for line-level defect prediction. To effectively demonstrate the statistical performance disparities between different approaches, we apply the Scott-Knott ESD test. Figures \ref{fig:fig3} and \ref{fig:fig4} present the Scott-Knott ESD rankings and the distributions of corresponding metrics for BAFLineDP and other advanced file-level and line-level defect prediction approaches within the WPDP scenario.

\begin{figure*}[t]
    \centering
    \subfigure[AUC ($\nearrow$)]{
        \label{fig:fig3:a}
        \includegraphics[width=0.315\textwidth]{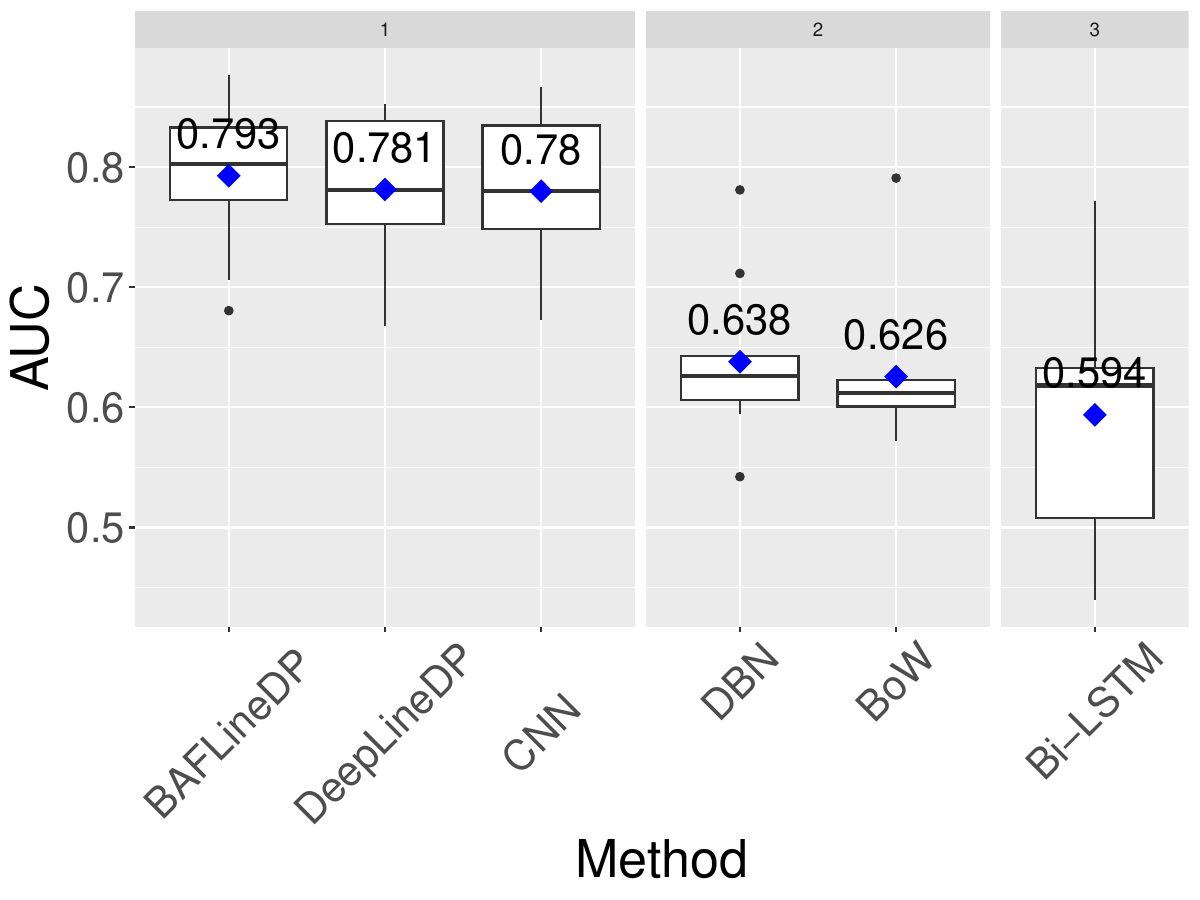}
    }
    \subfigure[BA ($\nearrow$)]{
        \label{fig:fig3:b}
        \includegraphics[width=0.315\textwidth]{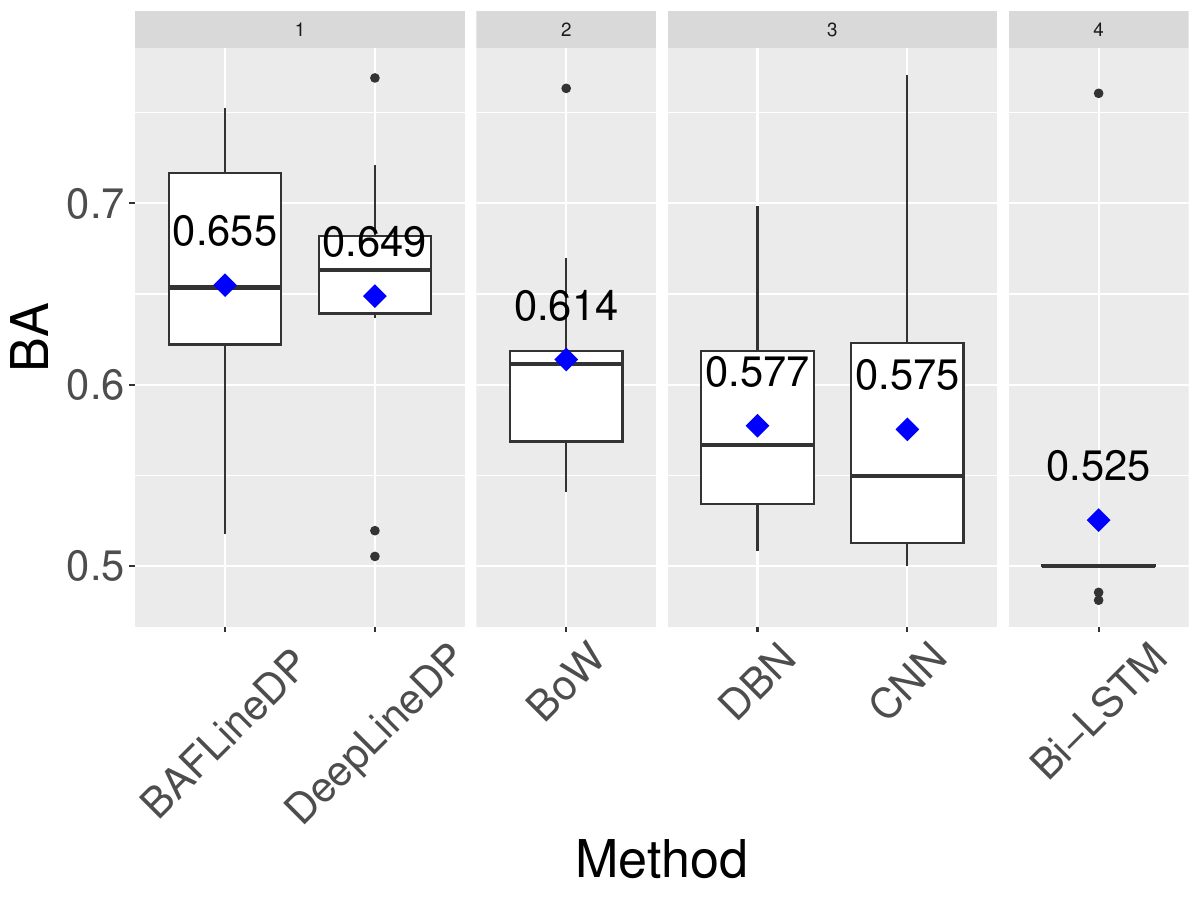}
    }
    \subfigure[MCC ($\nearrow$)]{
        \label{fig:fig3:c}
        \includegraphics[width=0.315\textwidth]{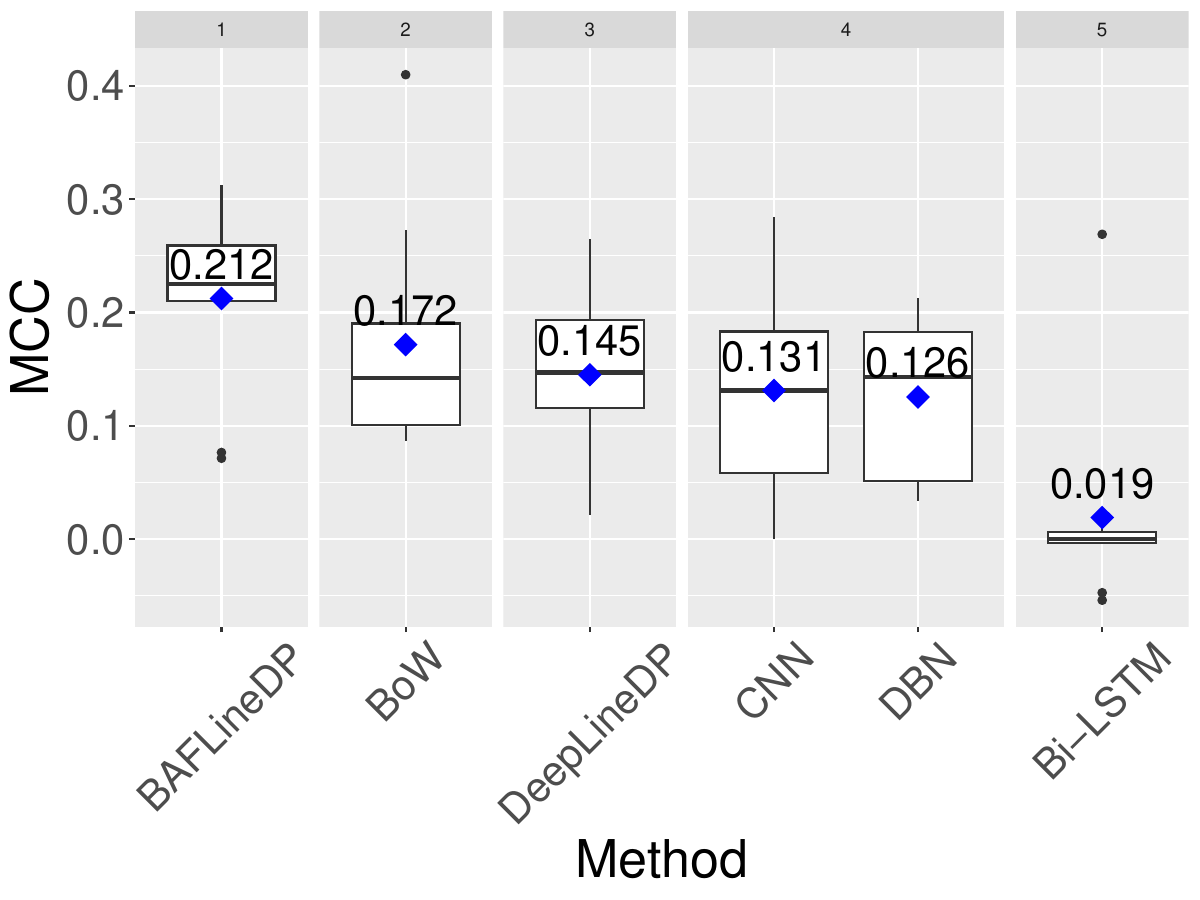}
    }
    \caption{(For RQ1) The Scott-Knott ESD rankings and the distributions of AUC, BA, and MCC of BAFLineDP and other file-level prediction approaches within the WPDP scenario. The higher ($\nearrow$) the values are, the better the approach is.}
    \label{fig:fig3}
\end{figure*}

\begin{figure}
    \centering
    \subfigure[Recall@Top20\%LOC ($\nearrow$)]{
        \label{fig:fig4:a}
        \includegraphics[width=0.315\textwidth]{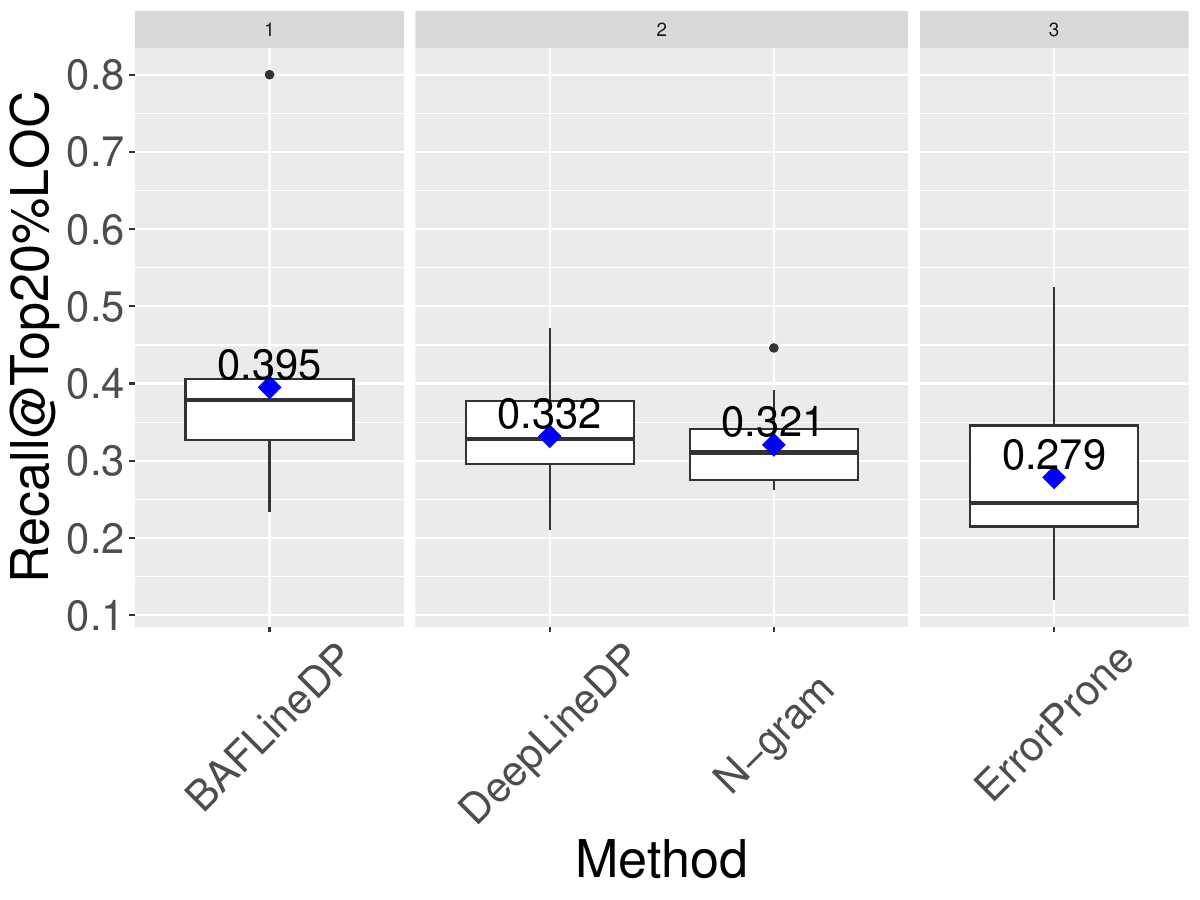}
    }
    \subfigure[Effort@Top20\%Recall ($\searrow$)]{
        \label{fig:fig4:b}
        \includegraphics[width=0.315\textwidth]{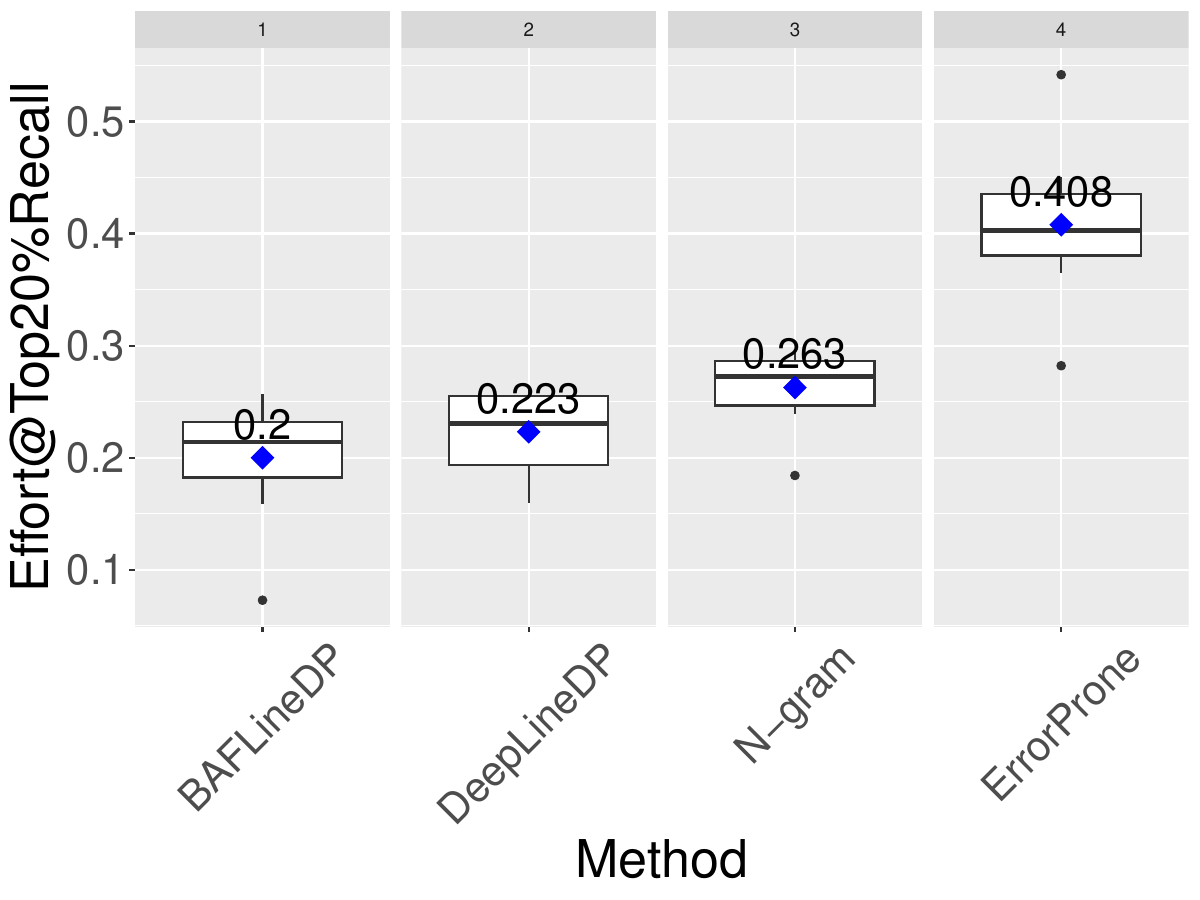}
    }
    \caption{(For RQ1) The Scott-Knott ESD rankings and the distributions of Recall@Top20\%LOC and Effort@Top20\%Recall of BAFLineDP and other line-level prediction approaches within the WPDP scenario. The higher ($\nearrow$) or the lower ($\searrow$) the values are, the better the approach is.}
    \label{fig:fig4}
\end{figure}

\uline{Results.} In light of the findings presented in Figure \ref{fig:fig3}, our BAFLineDP method demonstrates notably superior performance, as evidenced by its average AUC, BA, and MCC values, which respectively stand at 0.793, 0.655, and 0.212. These values exhibit a significant advantage, ranging from 2\% to 34\%, 1\% to 25\%, and 23\% to 1016\% when compared to those of other file-level defect prediction approaches. The outcome indicates the superiority of our BAFLineDP over existing file-level defect prediction methods. In addition, the Scott-Knott ESD test also confirmed that BAFLineDP consistently ranks among the best in terms of AUC, BA, and MCC, which shows that the performance difference is statistically significant with a non-negligible effect size. 

Notably, while BAFLineDP's median BA is slightly lower than DeepLineDP's, BA mainly assesses the precise differentiation between correct and defective files. However, in real-world software development, developers often favor a probabilistic ranking of defective files over a binary classification at a fixed threshold (e.g., 0.5). Therefore, when defective files are ranked based on probabilities, the superior performance of BAFLineDP is further confirmed by AUC, which assesses the ability to discriminate between defective and clean files.

Based on the outcomes illustrated in Figure \ref{fig:fig4}, our BAFLineDP method exhibits an average Recall@Top20\%LOC value of 0.395 and an average Effort@Top20\%Recall value of 0.2. Compared to other line-level defect prediction approaches, BAFLineDP demonstrates a notable enhancement in cost-effectiveness, boasting an improvement of 19\%-42\% in accurately identifying defective code lines with a fixed 20\% of the overall effort. Moreover, from a cost perspective, there is a reduction ranging from 10\% to 51\%. These findings firmly establish the superiority of BAFLineDP over existing line-level defect prediction approaches. Furthermore, the Scott-Knott ESD test also affirms that BAFLineDP consistently ranks among the top performers in terms of Recall@Top20\%LOC and Effort@Top20\%Recall, which shows that the performance difference is statistically significant with non-negligible effect size.

\begin{center}
    \fbox{\parbox{0.97\linewidth}{
    \textit{\textbf{Answer to RQ1:} The experimental findings within the WPDP scenario illustrate the superior performance of BAFLineDP in both file-level and line-level defect prediction compared to existing defect prediction approaches, achieving better cost-effectiveness and lower cost overhead.}
    }}
\end{center}

\begin{figure*}[t]
    \centering
    \subfigure[AUC ($\nearrow$)]{
        \label{fig:fig5:a}
        \includegraphics[width=0.315\textwidth]{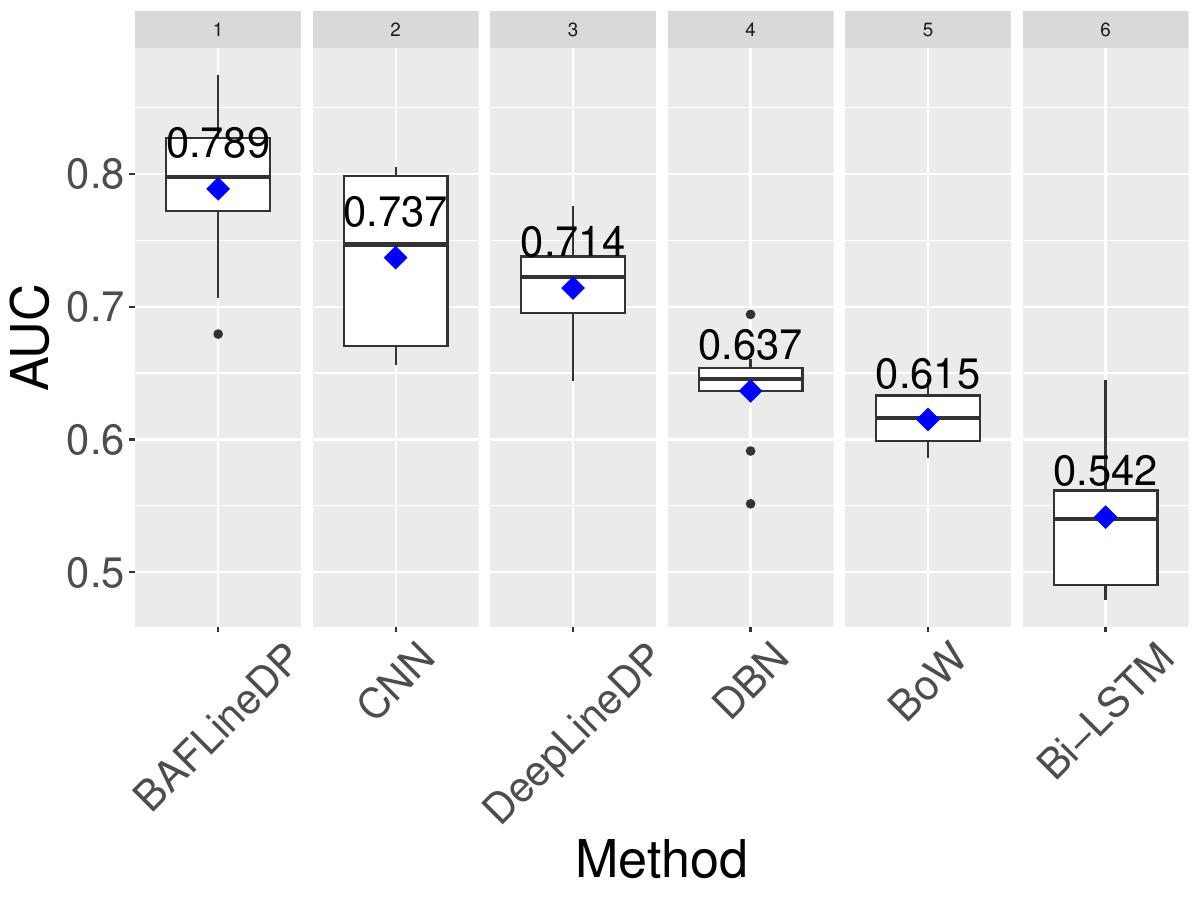}
    }
    \subfigure[BA ($\nearrow$)]{
        \label{fig:fig5:b}
        \includegraphics[width=0.315\textwidth]{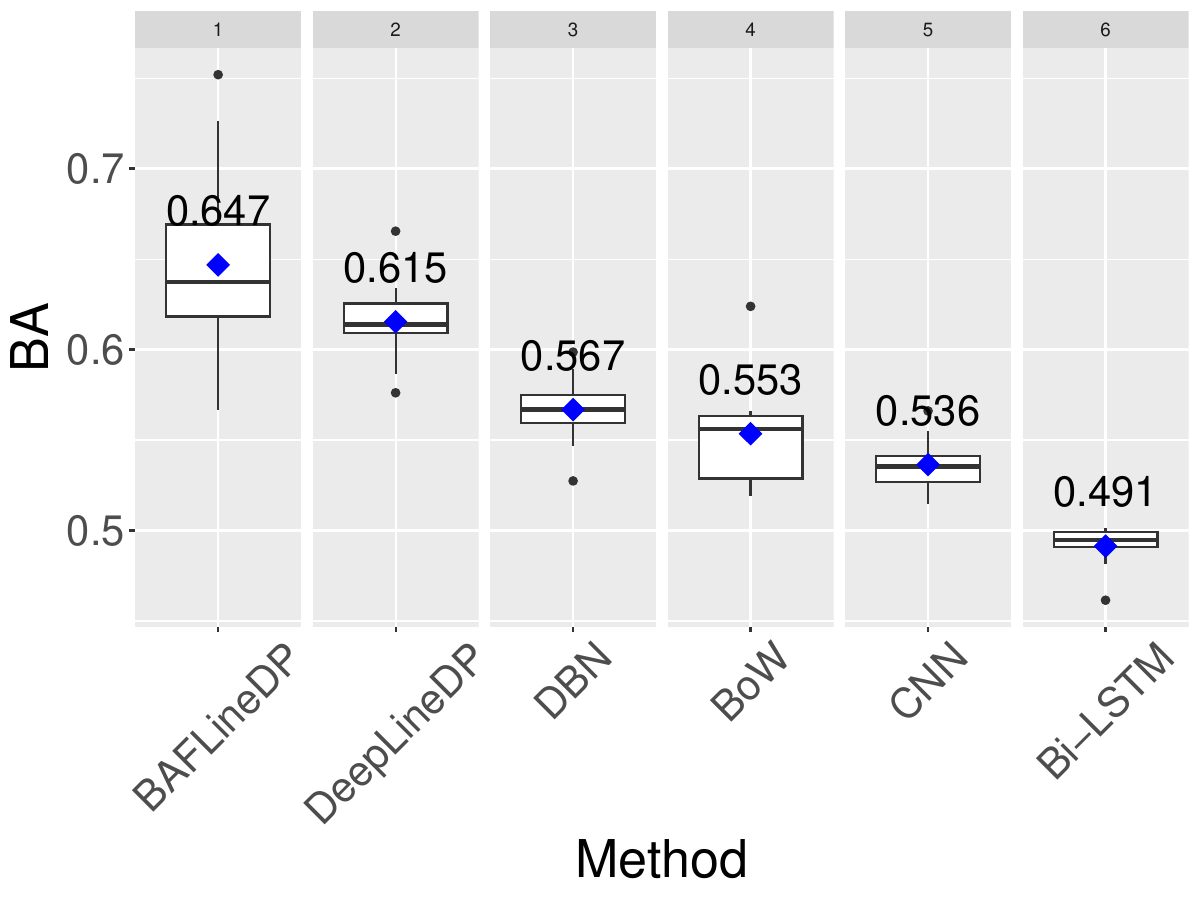}
    }
    \subfigure[MCC ($\nearrow$)]{
        \label{fig:fig5:c}
        \includegraphics[width=0.315\textwidth]{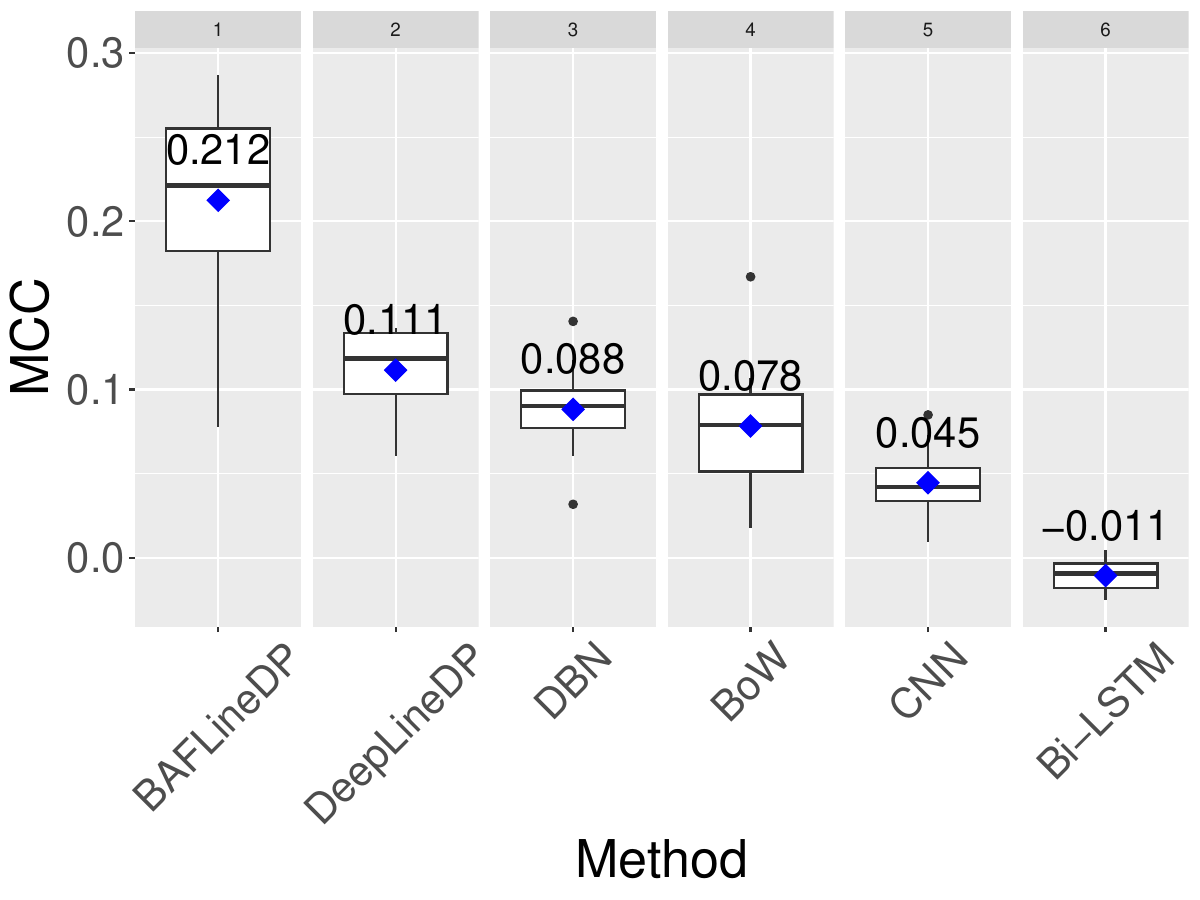}
    }
    \caption{(For RQ2) The Scott-Knott ESD ranking and the distributions of AUC, BA, and MCC of BAFLineDP and other file-level prediction approaches within the CPDP scenario. The higher ($\nearrow$) the values are, the better the approach is.}
    \label{fig:fig5}
\end{figure*}

\begin{figure}
    \centering
    \subfigure[Recall@Top20\%LOC ($\nearrow$)]{
        \label{fig:fig6:a}
        \includegraphics[width=0.315\textwidth]{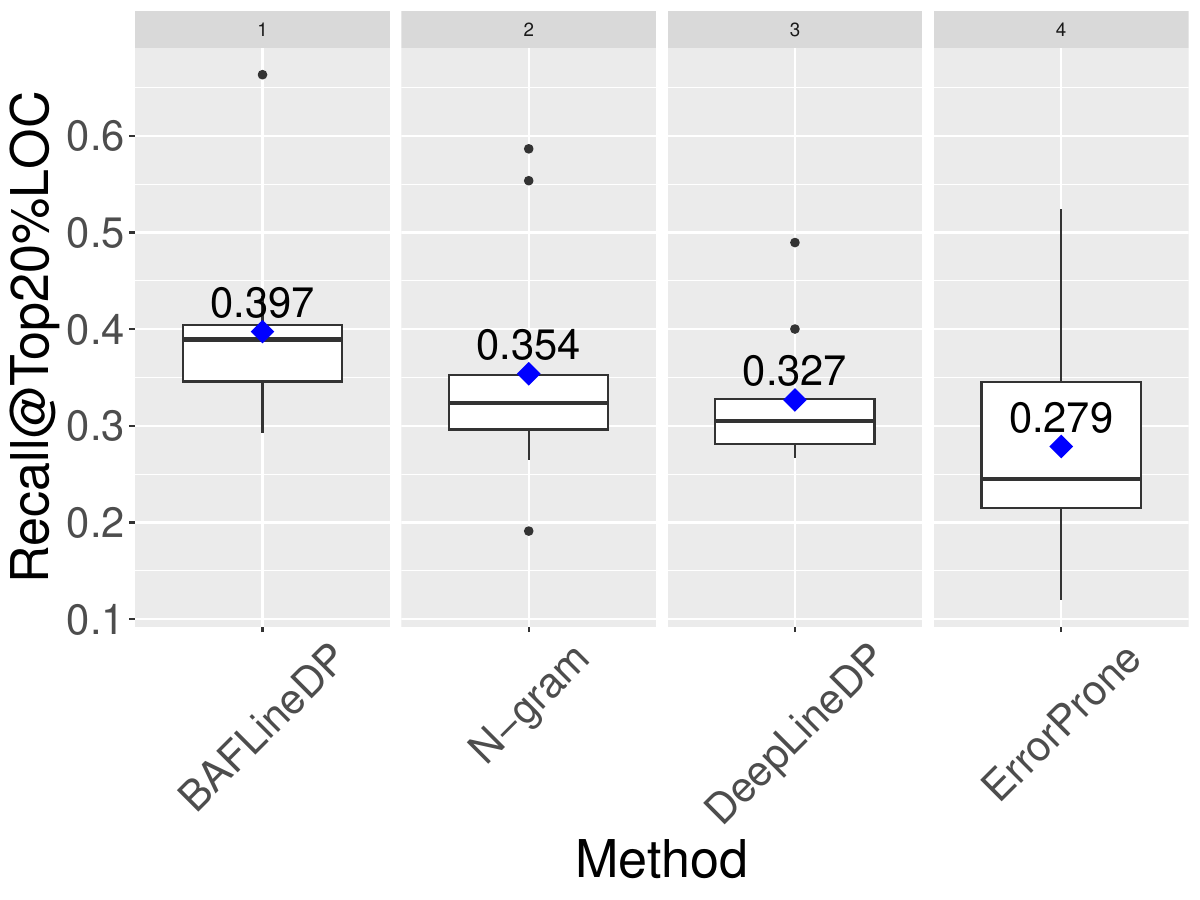}
    }
    \subfigure[Effort@Top20\%Recall ($\searrow$)]{
        \label{fig:fig6:b}
        \includegraphics[width=0.315\textwidth]{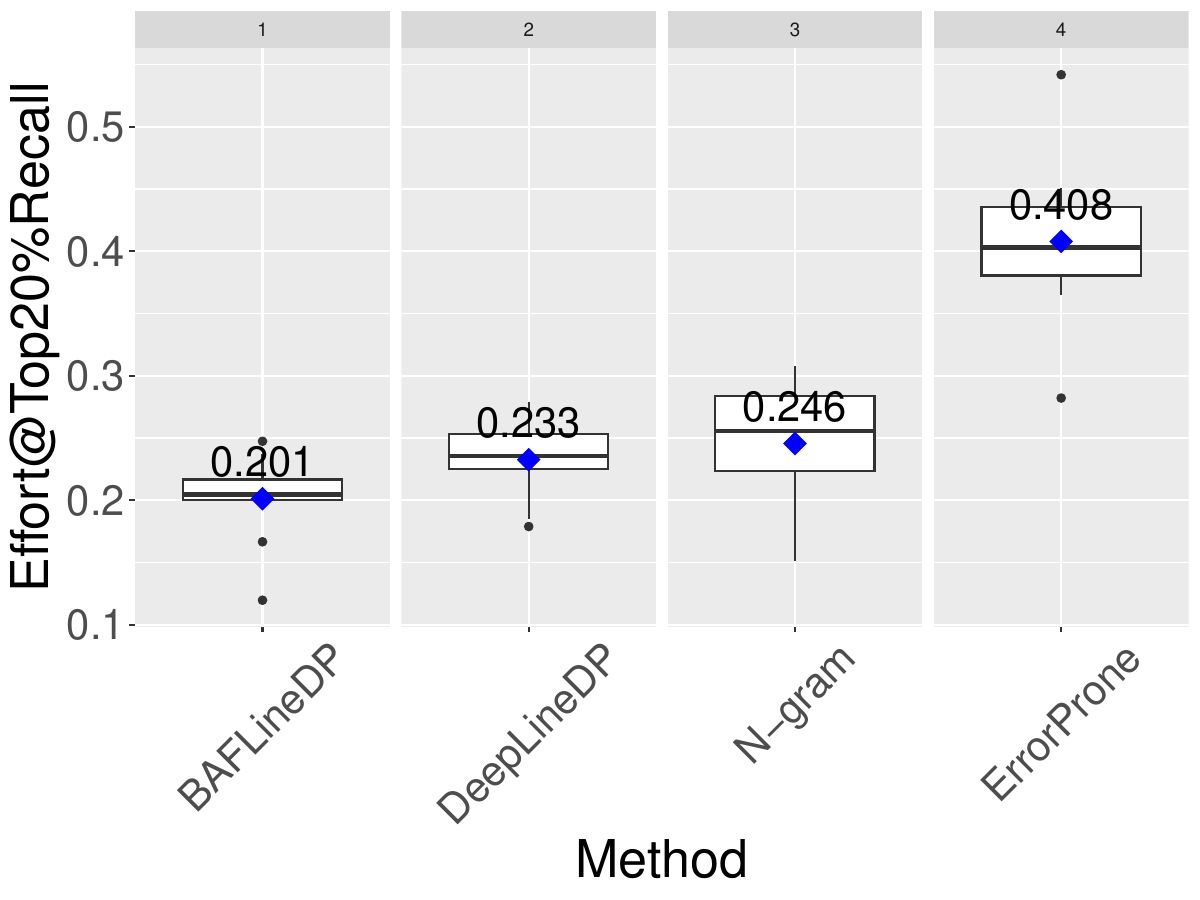}
    }
    \caption{(For RQ2) The Scott-Knott ESD ranking and the distributions of Recall@Top20\%LOC and Effort@Top20\%Recall of other line-level prediction approaches within the CPDP scenario. The higher ($\nearrow$) or the lower ($\searrow$) the values are, the better the approach is.}
    \label{fig:fig6}
\end{figure}

\textbf{\textit{(RQ2) How does BAFLineDP perform in both performance and cost-effectiveness within the CPDP scenario?}}

\uline{Motivation.} The majority of existing defect prediction methods are primarily assessed within the WPDP scenario, yielding favorable performance. However, practical applications often encounter challenges such as limited availability of training samples for new projects and substantial variations in project structures and complexities~\cite{jin2021cross}. Thus, although certain approaches excel in the WPDP scenario, their performance may not translate equally well to the CPDP scenario. Consequently, we focus on evaluating the performance and cost-effectiveness of BAFLineDP within the CPDP scenario while conducting a comparative analysis against other advanced file-level and line-level defect prediction methods.

\uline{Approach.} To address this inquiry, we selected 112 training-validation-testing within the CPDP scenario. For instance, we utilize Hive-0.9.0 and Hive-0.10.0 for training and verification, respectively, while testing is conducted on JRuby-1.5.0. Similar to WPDP, we evaluate the performance of BAFLineDP in file-level and line-level defect prediction using three traditional metrics (i.e., AUC, BA, and MCC) and two effort-aware metrics (Recall@Top20\%LOC and Effort@Top20\%Recall). Furthermore, we compared BAFLineDP against four file-level defect prediction methods (i.e., DBN, CNN, BoW, and Bi-LSTM) and three line-level defect prediction approaches (i.e., DeepLineDP, ErrorProne, and N-gram).

It is important to note that ErrorProne, an open-source static code analysis plug-in, is unaffected by training data in line-level defect prediction. Its performance solely relies on customized code rules. Consequently, the performance results of ErrorProne in CPDP line-level defect prediction align with those observed in WPDP line-level defect prediction.

To provide a comprehensive and statistically significant analysis of performance differences among the approaches, we again employ the Scott-Knott ESD test. Figures \ref{fig:fig5} and \ref{fig:fig6} illustrate the Scott-Knott ESD rankings and the distributions of corresponding indicators for BAFLineDP and other file-level and line-level defect prediction methods within the CPDP scenario.

\uline{Results.} According to the results in Figure \ref{fig:fig5}, the average AUC, BA, and MCC values of BAFLineDP are 0.789, 0.647, and 0.212, respectively, outperforming other file-level defect prediction methods by 7\%-46\%, 5\%-32\%, and 91\%-2027\%. These results establish BAFLineDP's superiority in the CPDP scenario. Additionally, the Scott-Knott ESD test confirmed BAFLineDP's consistent top performers in AUC, BA, and MCC, which indicates that the performance difference is statistically significant with a non-negligible effect size.

As shown in Figure \ref{fig:fig6}, the average Recall@Top20\%LOC and Effort@Top20\%Recall of BAFLineDP are 0.397 and 0.201, respectively. Compared to other line-level defect prediction methods, BAFLineDP exhibits a 12\% to 42\% improvement in identifying defective code lines using only 20\% of the overall effort and a 14\% to 51\% reduction in costs. These findings position BAFLineDP as a superior choice over existing line-level defect prediction methods within the CPDP scenario. Furthermore, the Scott-Knott ESD test further confirms BAFLineDP's top performance in Recall@Top20\%LOC and Effort@Top20\%Recall, indicating a statistically significant difference with non-negligible effect sizes.

\begin{center}
    \fbox{\parbox{0.97\linewidth}{
    \textit{\textbf{Answer to RQ2:} The experimental results illustrate that within the CPDP scenario, BAFLineDP outperforms existing defect prediction methods in both file-level and line-level defect prediction, while exhibiting higher cost-effectiveness and incurring lower cost overhead. These outcomes further reaffirm the effectiveness of BAFLineDP.}
    }}
\end{center}

\begin{table*}[htbp]
	\centering
	\renewcommand\arraystretch{1.2}
	\caption{Ablation Study results of BAFLineDP}
    \tabcolsep=0.35cm
	\begin{tabular}{c|c|c|c|c|c|c|c|c}
		\toprule
        \multirow{3}{*}{\textbf{Approach}} &
        \multicolumn{4}{c|}{\textbf{WPDP (Line-Level)}} & 
        \multicolumn{4}{c}{\textbf{CPDP (Line-Level)}} \\

        \cline{2-9}

        ~ & 
        \multicolumn{2}{c|}{\textbf{Recall@Top20\%LOC($\nearrow$)}} & 
        \multicolumn{2}{c|}{\textbf{Effort@Top20\%Recall($\searrow$)}} & 
        \multicolumn{2}{c|}{\textbf{Recall@Top20\%LOC($\nearrow$)}} & 
        \multicolumn{2}{c}{\textbf{Effort@Top20\%Recall($\searrow$)}} \\

        \cline{2-9}
        
		~ & \textbf{Abl.} & \textbf{Diff.} & \textbf{Abl.} & \textbf{Diff.} & \textbf{Abl.} & \textbf{Diff.} & \textbf{Abl.} & \textbf{Diff.} \\
		\midrule
		w/o CodeBERT & 0.346 & -12.4\% & 0.222 & +11.0\% & 0.336 & -15.4\% & 0.229 & +13.9\% \\
		w/o Bi-GRU & 0.365 & -7.6\% & 0.211 & +5.5\% & 0.382 & -3.8\% & 0.203 & +1\% \\
        w/o BAFN & 0.363 & -8.1\% & 0.255 & +27.5\% & 0.343 & -13.6\% & 0.260 & +29.4\% \\
        \midrule
        BAFLineDP & \textbf{0.395} & - & \textbf{0.200} & - & \textbf{0.397} & - & \textbf{0.201} & - \\
		\bottomrule
	\end{tabular}
	\label{table2}
\end{table*}

\section{Discussion}
\subsection{Effects of the Pivotal Modules within the BAFLineDP}
We conducted an ablation study with the principal objective of dissecting the individual contributions of pivotal modules within the BAFLineDP framework, discerning their respective impacts on the overall efficacy. Our scrutiny is particularly centered on the examination of three pivotal components, namely CodeBERT, Bi-GRU, and BAFN, given our conviction that they assume a pivotal role in the capture of essential facets pertaining to source code line defect semantics, global line-level context, and local interaction information of code lines. Furthermore, we chose to focus our ablation study on line-level defect prediction, as it holds greater practical applicability compared to file-level defect prediction.

Within the context of our ablation study, we observe the changes in the average Recall@Top20\%LOC and Effort@Top20\%Recall metrics resulting from the elimination or replacement of specific components, both in the WPDP and CPDP scenarios. The results of this ablation study are tabulated in Table \ref{table2}, wherein Abl. represents the performance of the simplified iteration of BAFLineDP, while Diff. signifies the performance version in comparison to BAFLineDP.

\textit{\textbf{Replacing CodeBERT with Doc2Vec:}} In our investigation of CodeBERT's impact, we substitute it with Doc2Vec~\cite{le2014distributed}, an unsupervised algorithm adept at acquiring fixed-length feature representations from variable-length textual data, for encoding lines of code. As demonstrated in Table \ref{table2}, after replacing CodeBERT, the cost-effectiveness of line-level defect prediction in WPDP and CPDP dropped by 12.4\% and 15.4\%, respectively, while the cost overhead increased by 11.0\% and 13.9\%, respectively.

\textit{\textbf{Removing Bi-GRU:}} In exploring the role of Bi-GRU, we eliminate the utilization of Bi-GRU, thereby discontinuing the extraction of contextual information from code lines. As presented in Table \ref{table2}, the removal of Bi-GRU led to a decline in cost-effectiveness for line-level defect prediction in both WPDP and CPDP scenarios, registering reductions of 7.6\% and 3.8\%, respectively, while the cost overhead experienced a rise of 5.5\% in WPDP and 1\% in CPDP.

\textit{\textbf{Removing BAFN:}} To investigate the significance of BAFN, the component responsible for local interaction information between source code lines and their contextual counterparts, we remove BAFN. As depicted in Table \ref{table2}, the removal of BAFN resulted in a marked reduction in cost-effectiveness for line-level defect prediction in WPDP and CPDP scenarios, with drops of 8.1\% and 13.6\%, respectively. In parallel, the cost overhead exhibited a substantial escalation of 27.5\% in WPDP and 29.4\% in CPDP.

In summary, CodeBERT, Bi-GRU, and BAFN each play pivotal roles in capturing essential elements of source code line defect semantics, global line-level context, and local interaction information of code lines. This contribution significantly enhances the cost-effectiveness of SQA resource management, concurrently alleviating the burden on developers by reducing the need for manual inspection of defective code lines and the associated workload.

\subsection{Threats To Validity}
In this section, we describe several threats that may have an impact on the effectiveness of our approach.

\subsubsection{\textbf{Implementation of comparison methods}}
We implemented benchmark methods such as DeepLineDP and ErrorProne using open-source code to reduce the possible impact of incorrect implementations. For methods that do not provide source code (e.g., CNN, Bi-LSTM, BoW, DBN, and N-gram), we strictly follow the implementation details in the relevant papers, but there may still be some deviations.

\subsubsection{\textbf{The experimental results may not be generalizable}}
We conducted experiments on 9 open-source software projects with different sizes and defect rates. It would be beneficial to generalize our research while avoiding the specificity of experimental results. However, we cannot guarantee that our method will achieve the same improvement on other software datasets.

\subsubsection{\textbf{The model hyperparameter selection does not consider all options}}
In our experiments, we try to adjust the hyperparameters of the BAFLineDP model to obtain better defect prediction performance. However, it is impractical to evaluate all possible hyperparameter combinations. We evaluated several hyperparameter combinations within specific ranges based on previous research experience~\cite{pornprasit2022deeplinedp}.

\section{Conclusion}
In this paper, we introduce BAFLineDP, a novel line-level defect prediction approach based on a code bilinear attention fusion framework. This methodology can effectively merge source code line semantics, line-level context, and local interaction information between code lines and corresponding line-level context to identify defective code files and lines. An empirical study on within- and cross-project defect prediction on 9 software projects covering 32 versions demonstrates that BAFLineDP outperforms existing file-level and line-level defect prediction methods. Thus, we anticipate that BAFLineDP can help software quality assurance teams find defective lines of code in a cost-effective manner. The data and source code that support the findings of this study are openly available on GitHub at https://github.com/insoft-lab/BAFLineDP.

\section*{Acknowledgment}
This study was funded partly by Guangdong Natural Science Fund Project (2022A1515110564), Guangzhou Science and Technology Plan Project (202201010312), Key Research and Development Plan of Guangzhou (202206010091, 2023B03J1363), Special Fund for Rural Revitalization Strategy of Guangdong (2023TS-3), Science and Technology Project of Meizhou City Tobacco Monopoly (202304).

\normalem   
\bibliographystyle{IEEEtran}
\bibliography{reference.bib}

\end{document}